\RequirePackage{fix-cm}
\documentclass[twocolumn,epjc3]{svjour3}
\smartqed
\RequirePackage{graphicx}
\usepackage{cite, amssymb, amsmath}
\journalname{Eur. Phys. J. C}

\begin{document}

\title{Gravitational Field of a Spherical Perfect Fluid}
\author{Merab Gogberashvili\thanksref{e1,addr1,addr2}
        \and
        Beka Modrekiladze\thanksref{e2,addr3}}
\thankstext{e1}{e-mail: gogber@gmail.com}
\thankstext{e2}{e-mail: b.modrekiladze@gmail.com}

\institute{Javakhishvili Tbilisi State University, 3 Chavchavadze Avenue, Tbilisi 0179, Georgia \label{addr1}
           \and
           Andronikashvili Institute of Physics, 6 Tamarashvili Street, Tbilisi 0177, Georgia \label{addr2}
           \and
           Department of Physics, Carnegie Mellon University, Pittsburgh, PA 15213, USA \label{addr3}
}

\date{Received: date / Accepted: date}

\maketitle

\begin{abstract}
Analyzing the spacetime for a static spherically distributed perfect fluid we show that the smooth matching of the interior and exterior metrics for a realistic source is possible only for the distances from the origin that exceeds the photon sphere radius for this object.
\end{abstract}


The most important solution to the Einstein equations is the static spherically symmetric metric, which in the Schwarzschild gauge can be written in the form \cite{Weyl}:
\begin{eqnarray} \label{metric}
ds^2 &=& \left[1 - \frac {A(r)}{r}\right] y^2(r) dt^2 - \frac {1}{\left[1-A(r)/r\right]} dr^2 - \nonumber \\
&-& r^2 d\theta^2 - r^2 \sin^2 \theta d\phi^2~.
\end{eqnarray}
The system of Einstein equations for this \emph{ansatz} is:
\begin{eqnarray} \label{Einstein}
8\pi T^t_t &=& \frac {A'}{r^2} ~, \nonumber \\
8\pi T^r_r &=& \frac{A'}{r^2} -2 \left(r - A\right)\frac{y'}{r^2y} ~, \\
8\pi T^\theta_\theta = 8\pi T^\phi_\phi &=& - \frac{(r-A)y''}{ry} + \frac {A''}{2r} + \nonumber \\
&+& \frac{y'}{r^2y}\left(\frac {3rA'}{2}- \frac {A}{2} - r\right)~,\nonumber
\end{eqnarray}
where primes denote derivatives with respect to the radial coordinate $r$.

In the Schwarzschild radial gauge, from the first equation of this system one can find
\begin{equation} \label{A=2GM}
A(R) = 8\pi \int_0^R dr\, r^2 T^t_t ~,
\end{equation}
which is regarded to be a measure of the amount of mass located within a surface of the radius $R$. To the best of the knowledge, no corresponding study has ever been made for other choices of radial gauge and there is no known formula for mass in other coordinates. For simplicity in (\ref{A=2GM}) we set $A(0)=0$, since in this paper we don't touch the central singularity problem; also we want to consider the matching region with a Schwarzschild exterior, were $A$ should reduce to a positive constant.

Most isolated bodies, large enough to support strong gravitational fields, are constructed of material in a state of high fluidity and the most natural assumption for the spatial components of the static energy-momentum tensor in (\ref{Einstein}) would be the equation of state of a perfect fluid:
\begin{equation} \label{T=T=T}
T^r_r = T^\theta_\theta = T^\phi_\phi = - p(r)~,
\end{equation}
where $p(r)$ represents the pressure, which usually is assumed to be a positive quantity. The relation (\ref{T=T=T}) leads to the second order non-linear differential equation for the metric functions $A(r)$ and $y(r)$:
\begin{equation} \label{y}
2r(r - A)y'' + (5A - 3rA' - 2r)y' + (2A' - r A'')y = 0~.
\end{equation}
Boundary conditions for (\ref{y}) are:
\begin{equation} \label{A_s}
A(r)|_{r>R} \to 2M ~, \qquad \qquad y(r)|_{r>R} \to 1~,
\end{equation}
which follow from the requirement to have the vacuum Schwarzschild metric outside the surface of the junction $r=R$.

We will consider the solutions of (\ref{y}) for interiors of stellar objects and an important role in the determination of their physical character is played by the way in which they can be matched with a Schwarzschild exterior (\ref{A_s}). Besides the vacuum spherically symmetric spacetime (\ref{A_s}), the well-known solution of (\ref{y}) corresponds to the Schwarzschild metric for a static sphere of perfect fluid \cite{Sch},
\begin{equation}
A(r) = \frac {2M}{R^3}r^3~, \qquad y(r) = \frac 52 -\frac 32 \sqrt{\frac{R^3 - 2MR^2}{R^3 - 2Mr^2}}~,
\end{equation}
which can match to the exterior solution with (\ref{A_s}) across the surface of a radius $R \geq 9M/4$, on which the pressure goes to zero \cite{Buch-1}. However, this model of homogeneous static sphere of perfect fluid implies too much mass inside the radius $R$ and too little outside and contains no provisions for continuity, a characteristic which nature deems almost necessary for any realistic spherical sources. Also, the obtained junction radii, $R \geq 9M/4$ \cite{Buch-1} (or $R \geq 5M/2$ \cite{Buch-2}) allow matching of the internal and external Schwarzschild's solutions even inside the photon sphere $R=3M$. However, the photon sphere is the lower bound for any stable orbit and it seems that such static object, similar to a black hole, cannot be realistic.

Since Schwarzschild's paper \cite{Sch}, little progress has been made toward other analytic solutions of Einstein equations for realistic mass distributions, most recent attempts have dealt with various methods of approximations. Admitted model for the structure of relativistic stellar objects is unknown. Perhaps the most satisfactory approach to obtain an accurate solution of (\ref{y}) in the junction region, $r\sim R$, is to expand the internal density in a series of positive powers of $r$, and the external density in a series of negative powers of $r$.

Let us introduce some general conditions for the physically reasonable metric function $A(r)$ close to the junction surface $r = R$. A solution will correspond to actual astrophysical objects only if the following conditions are satisfied \cite{Del-Lake}:
\begin{enumerate}
\item Density and pressure should be positive inside the object;
\item The gradients of density and pressure should be negative;
\item The speed of sound should be less than the speed of light;
\item The energy conditions should be satisfied.
\end{enumerate}

First of all, it is clear that for smooth junction with the vacuum Schwarzschild metric (\ref{A_s}), the function $A(r)$ should be continuous (the mass density $T_t^t$ is assumed to be at least sectionally continuous) and positive (at least in the junction region $r \to R$,). Since density $T_t^t$ is assumed to be positive, from the first equation of the system (\ref{Einstein}) we see that $A'(r)$ should be also positive. Thus, the first out of above four requirements on a physically reasonable solution, in our gauge (\ref{metric}) gives the conditions just on the single metric function,
\begin{equation} \label{A>0}
A(r) > 0~, \qquad A'(r) > 0~.
\end{equation}

Regardless of the mass distribution, at large distances from the center of symmetry, the energy density $T_t^t$ must fall off faster than the inverse cube in order that the total mass be finite. Thus, from the first equation of the system (\ref{Einstein}) it is clear that close to a junction surface $A'(r)$ should be decreasing function, i.e. the second above requirement on a physically reasonable solution gives:
\begin{equation} \label{A''<0}
A'' < 0~, \qquad \left(rA'\right)'< 0~.
\end{equation}

Remarkably, very few of all known exact spherically symmetric solutions of the Einstein equations actually satisfy the conditions (\ref{A>0}) and (\ref{A''<0}). The examples of some exceptions are:
\begin{itemize}
\item The solution III of Bayin, see (2.22) in \cite{Bayin},
\begin{equation}
A_{\rm III}(r) = r\left(1 + C_0 - \frac{2}{r^2} \right)~,
\end{equation}
when the integration constant $C_0 > 2/R^2 - 1$;
\item The solution IV from the same Bayin paper, (2.26) in \cite{Bayin},
\begin{equation}
A_{\rm IV}(r) = r\left(1 - \frac{3C_0 + C_1 r^2 - 2 r^3}{4^{2/3}(3C_0 + 4r^3)} \right)~,
\end{equation}
when the integration constant $C_1$ equals to zero and $C_0 > 2R^3/3$;
\item The solution (89) in the Carloni and Vernieri paper \cite{Car-Ver},
\begin{equation}
A_{C-V} (r) = \frac{4 r^3}{C_0 + 8r^2}~,
\end{equation}
when $0 < 3C_0< 8 R^2$.
\end{itemize}

Let us now estimate the distance where the function $A(r)$ for a reasonable mass distribution can be considered as a constant and reduce to the Schwarzschild's vacuum solution (\ref{A_s}). Using (\ref{Einstein}) and (\ref{T=T=T}), from the requirement to have non-negative pressure $p(r) \ge 0$, we find:
\begin{equation} \label{A'/r<}
\frac{A'}{r} \leq 2\left(1 - \frac {A}{r}\right)\frac{y'}{y} ~.
\end{equation}
On the other hand, from the dominant energy condition $T^t_t \geq |T^r_r|$, we have:
\begin{equation} \label{A'/r>}
\frac{A'}{r} \geq \left(1 - \frac {A}{r}\right)\frac{y'}{y} ~.
\end{equation}
So close to the junction surface, $r \sim R$, we can assume
\begin{equation} \label{A'/r=}
\frac{A'}{r-A} = 2a \frac{y'}{y} ~,
\end{equation}
where $a$ is a constant given by:
\begin{equation} \label{a}
\frac 12 \leq a \leq 1~.
\end{equation}
When $a = 1$ in (\ref{A'/r=}) the equation (\ref{y}) has the solution $y'= 0$. Therefore, from $a \to 1$ it follows that $y' \to 0$ and thus $A' \to 0$, which corresponds to the smooth junction of the interior metric tensor and its first derivatives with the vacuum Schwarzschild metric (\ref{A_s}).

The relation (\ref{A'/r=}) brings the equation (\ref{y}) to the form:
\begin{equation} \label{y''=}
r(r - A)y'' = \left[r(1 + A') - \frac {5-4a}{2(1-a)}A\right]y'~.
\end{equation}
Using this equation and the estimation (\ref{A'/r=}), the first inequality in (\ref{A''<0}) takes the form:
\begin{equation} \label{A' + rA''}
A'' = 2a \left\{\left[2 - \frac {5-4a}{2(1-a)}\frac Ar \right]\frac{y'}{y} - (r-A)\frac{y'^2}{y^2} \right\}< 0~.
\end{equation}
Close to the junction surface, $A' \to const$. Therefore, $A'' \to 0$ and from (\ref{A' + rA''}) it follows that:
\begin{equation} \label{y'/y>}
\frac{y'}{y} \gtrsim \frac {2r - \frac {5-4a}{2(1-a)}A}{r(r-A)}~.
\end{equation}
Since due to (\ref{A>0}) $A, A' > 0$ and according to (\ref{A'/r=}) the ratio $y'/y$ should be also positive, then (\ref{y'/y>}) gives us the condition:
\begin{equation} \label{r>}
r > \frac {5-4a}{4(1-a)}A~.
\end{equation}
For the lowest possible value of the constant $a$ from (\ref{a}), for the acceptable value of the junction radius we have
\begin{equation}
R > \frac 32 A \sim 3M~,
\end{equation}
which exceeds the photon sphere, $R=3M$, and is larger than the estimates obtained in \cite{Buch-1, Buch-2}. From  (\ref{r>}) it is clear that for more realistic values of the constant $a \to 1$, which corresponds to the smooth junction of the metric tensors and its first derivatives, the acceptable junction radius should be much larger.

To conclude, by analyses of the interior and exterior Schwarzschild metrics for a spherically distributed perfect fluid with the reasonable equation of state in the junction region, we have demonstrated that in physically relevant cases these metrics can be continuously matched only outside the photon sphere of the object.


\begin{acknowledgements}
This work was supported by SRNSFG (Shota Rustaveli National Science Foundation of Georgia) [DI-18-335/New Theoretical Models for Dark Matter Exploration].
\end{acknowledgements}


\end{document}